\documentclass[secnumarabic,nobibnotes,aps,prl,superscriptaddress,floatfix]{revtex4}
\usepackage{amsmath}
\usepackage{amssymb}
\usepackage{graphicx}
\usepackage{subfigure}
\usepackage{epstopdf}
\usepackage{color}
\usepackage{ulem}
\usepackage{bm}

\newcommand{\YbTiO}{{\rm Yb_{2}Ti_{2}O_{7}}}

\newcommand{\CW}{\theta_{\rm CW}}

\newcommand{\crysvec}[1]{[#1]}

\newcommand{\Emu}{E_{\mu}}
\newcommand{\Enu}{E_{\nu}}

\begin{document}




\title{Rods of Neutron Scattering Intensity in Yb$_2$Ti$_2$O$_7$: Evincing a 
Hamiltonian with Significant Anisotropic Exchange in a Magnetic Pyrochlore Oxide}

\author{Jordan D. Thompson}
\author{Paul A. McClarty}
\affiliation{Department of Physics and Astronomy, University of Waterloo, Waterloo, ON, Canada, N2L 3G1}
\author{Henrik M. R\o{}nnow}
\affiliation{Laboratory for Quantum Magnetism, \'{E}cole Polytechnique F\'{e}d\'{e}rale de Lausanne (EPFL), 
CH-1015 Lausanne, Switzerland}
\author{Louis P. Regnault}
\affiliation {CEA-Grenoble, INAC-SPSMS-MDN, 17 rue des Marthyrs, 38054 Grenoble, cedex 9, France}
\author{Andreas Sorge}
\affiliation {Network Dynamics Group, MPI for Dynamics and Self-Organization, Bunsenstr. 10, 37073 G\"{o}ttingen, Germany}
\affiliation{Department of Physics and Astronomy, University of Canterbury, Private Bag 4800, Christchurch, New Zealand}
\author{Michel J. P. Gingras}
\affiliation{Department of Physics and Astronomy, University of Waterloo, Waterloo, ON, Canada, N2L 3G1}
\affiliation{Department of Physics and Astronomy, University of Canterbury, Private Bag 4800, Christchurch, New Zealand}
\affiliation{Canadian Institute for Advanced Research, 180 Dundas Street West, Toronto, Ontario M5G 1Z8, Canada}

\date{\today}

\begin{abstract}
Paramagnetic correlations in the 
magnetic material Yb$_2$Ti$_2$O$_7$ have been investigated
 via neutron scattering, revealing a $[111]$ rod of 
scattering intensity. Assuming interactions between the Yb$^{3+}$ 
ions composed of all symmetry-allowed nearest neighbor exchange 
interactions and long-range dipolar interactions, we construct
a model Hamiltonian that allows for an excellent description of the 
neutron scattering data.
Our results provide compelling evidence for significant 
anisotropic exchange interactions in an insulating magnetic pyrochlore oxide.
 We also compute the real space correlations leading to the $[111]$ rod of scattering.
\end{abstract}

\maketitle



In geometrically frustrated magnetic materials there 
exists no configuration of magnetic moments that 
simultaneously satisfies all the pairwise magnetic interactions. 
Experimental and theoretical research over the past twenty years has 
shown that frustrated magnetic systems are prone to exhibit
novel and intriguing collective thermodynamic phenomena \cite{Diep_Springer}.

Among frustrated three dimensional systems, the
{\it A}$_2${\it B}$_2${\rm O}$_7$ pyrochlores 
have attracted much attention \cite{GGG}. 
In these compounds, {\it A} is a trivalent rare earth ion 
(Ho, Dy, Tb, Gd, Yb) or yttrium (Y) 
and {\it B} is a tetravalent transition metal ion (Ti, Sn, Mo, Mn).
Both {\it A} and {\it B} reside on two distinct lattices
of corner-sharing
tetrahedra.
Theory predicts that classical~\cite{Moessner} and quantum~\cite{Canals}
Heisenberg spins on a pyrochlore lattice interacting via an isotropic
antiferromagnetic nearest neighbor exchange Hamiltonian, $H_{\rm H}$, 
fail to develop conventional  LRO down to zero temperature. 
In real pyrochlore compounds, however, there generally exists some
combination of other perturbing magnetic interactions (e.g.
single ion anisotropy, dipolar interactions, etc) beyond $H_{\rm H}$.
Since $H_{\rm H}$ alone does not produce LRO,
the low temperature magnetic correlations of these materials
are strongly influenced by the competition between material-specific perturbations. 
This is the origin of the richness of phenomena 
observed in the $A_2B_2O_7$ pyrochlores~\cite{GGG} including
spin liquid~\cite{Gardner-TTO-PRL}, spin glass~\cite{Gingras-YMoO-PRL}, 
spin ice~\cite{Bramwell-Science},
and LRO with persistent low-temperature spin dynamics~\cite{Gd2Sn2O7,111rod}.
In this article, we consider the Yb$_2$Ti$_2$O$_7$ pyrochlore
which does not apparently
exhibit any of the aforementioned phenomena and has some unique and unusual
features of its own which have heretofore remained unexplained.


Yb$_2$Ti$_2$O$_7$ has a ferromagnetic character with a 
Curie-Weiss temperature, 
$\theta_{\rm CW} = +0.65 \pm 0.15$ K \cite{Bramwell-JPC,Hodges-YbTO-JPC}.
 The Yb$^{3+}$ $\sim 3$ $\mu_{\rm B}$ magnetic moments predominantly lie
perpendicular to the local $[111]$ cubic unit cell diagonals, making
this system the only known local $[111]$ XY pyrochlore
with a ferromagnetic $\theta_{\rm CW}$~\cite{GGG}.
Magnetic specific heat ($C_m$) measurements reveal a sharp first order transition at 
$T_c\approx 240$ mK \cite{Blote}, suggesting the onset of LRO.
While a single crystal elastic neutron scattering (NS) study suggested
ferromagnetic order below $T_c$ \cite{Yasui},
a subsequent polarized NS study~\cite{Gardner-YbTO} did not confirm such ordering.
Furthermore, powder NS shows no LRO down to
$110$ mK \cite{Hodges-YbTO-PRL} and very recent NS 
on a single crystal sample has not found
any sign of LRO in a broad region of the (hkk) 
scattering plane at 30 mK \cite{Ross}.
The $T_c\approx 240$ mK transition seen in $C_m$ has therefore so far not been
matched with the observation of conventional (dipolar magnetic) LRO.
In addition, M\"ossbauer spectroscopy
and muon spin relaxation ($\mu$SR) measurements find a rapid decrease of the 
Yb$^{3+}$ magnetic moments fluctuation rate, $\nu$, upon approaching $T_c$ from above, with
$\mu$SR revealing a temperature-independent $\nu$ 
(i.e. persistent spin dynamics) from $T_c$ down to $40$ mK, the lowest
temperature considered~\cite{Hodges-YbTO-PRL}.
Considering all these results together, one may ask whether
the $240$ mK transition in Yb$_2$Ti$_2$O$_7$ 
may be another rare example of hidden (non-dipolar) order~\cite{Santini-RMP}.
Another intriguing possibility~\cite{Hodges-YbTO-PRL} 
is that the 240 mK first order transition 
takes place between a ``spin gas'' (paramagnetic) state
and a spin liquid without any symmetry breaking.

A very interesting feature of the magnetic correlations in 
Yb$_2$Ti$_2$O$_7$ found at temperatures $T_c < T \lesssim 2$ K are rods
of NS intensity along the $[111]$ directions \cite{111rod,Ross}.
At first sight, the presence of such rods signals an anisotropy in the magnetic correlations 
that may originate from a structural transition at $T\gtrsim 2$ K or from intrinsically
stronger correlations within the kagome planes perpendicular to the four $[111]$
directions forming the undistorted pyrochlore structure as compared to
 correlations perpendicular to the kagome planes \cite{Ross} $-$ hence making a ``spin liquid crystal'' of sorts.


Here, we report results from diffuse NS 
measurements on Yb$_2$Ti$_2$O$_7$  
in a temperature range above $\theta_{\rm CW}> T_c$. 
By numerically annealing a set of exchange couplings to maximize the agreement between
experimental results and NS computed within a random phase approximation (RPA), 
we determine a spin Hamiltonian, ${H}$, that captures the main features
 of the observed paramagnetic NS pattern and 
reveals that significant spin exchange 
anisotropy exists in this insulating pyrochlore oxide material.
There have been recent claims of evidence for
anisotropic exchange at play in
Yb$_2$Ti$_2$O$_7$ \cite{Cao,Bonville-JPC} and
other $A_2B_2{\rm O}_7$ pyrochlores \cite{Cao,Malkin}.
However, because of the limitations imposed by 
the physical quantity considered (local susceptibility), 
and the models used in these works, the specific
nature and symmetry of the putative microscopic exchange 
has, until this work remained hidden \cite{Thompson-JPCM}.
Owing to the highly structured spin correlation functions in
Yb$_2$Ti$_2$O$_7$, we obtain compelling evidence for
anisotropic exchange.
From the reciprocal space structure factor, we compute the
spin-spin correlations along different crystallographic directions.
While these correlations are anisotropic, the correlation 
lengths themselves do not distinguish between correlations 
parallel and perpendicular to the kagome planes.




The NS cross-section was measured on the D23
diffractometer 
at the Institut Laue Langevin, France. 
A single crystal rod was aligned with $[{\rm h}00]$ and $[0{\rm kk}]$ in the scattering plane, 
hence providing access to all principal symmetry
directions of the cubic crystal structure. With incident neutron energy of 14.7~meV, 
significantly larger than any characteristic
energy scale in the system, and counting all final energies, 
the measured intensity is proportional to the spatial Fourier
transform of the instantaneous correlation function 
$S(\mathbf{q}) = \int S(\mathbf{q},\omega) d\omega$ 

\begin{figure}[!htb]
\begin{center}
\includegraphics[width=0.5\columnwidth]{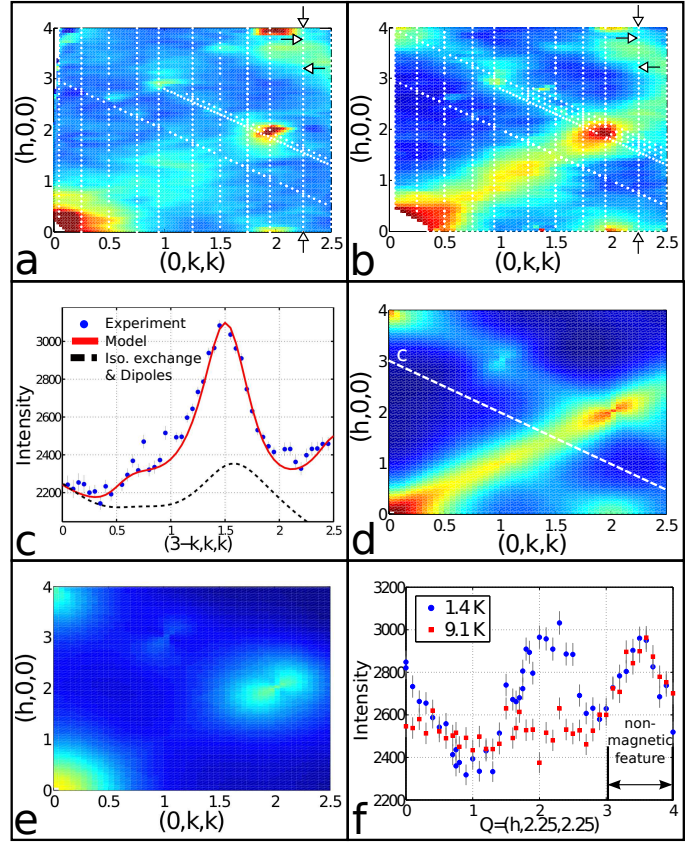}
\end{center}
\caption{(Color online) Neutron scattering (NS) maps in the (h,k,k) plane.
(a) and (b) show experimental data at 9.1 K and 1.4 K, respectively.
 In (b), a rod of scattering intensity along $\crysvec{111}$ is clearly seen,
while in (a) the rod is not strong. 
(c) shows a comparison between experimental data and the computed NS intensity at $1.4$ K along 
$\left[3-{\rm k},{\rm k},{\rm k} \right ]$
for both isotropic exchange and the full model.
 (d) shows computed NS using the Hamiltonian $H$ (see text),
 at T $=1.4$ K.  (e) shows the calculated NS using $H$ with 
only isotropic exchange determined from fitting $\CW$
and long-range dipolar interactions at T $=1.4$ K.
(f) shows experimental neutron scattering at $1.4$ K and $9.1$ K along the
line $\left[{\rm h},2.25,2.25\right]$ showing that the feature between h = $3$ and h = $4$
 does not change in intensity with temperature. The white arrows in (a) and (b)
indicate the range h $\in [3,4]$ in $\left [{\rm h},2.25,2.25\right]$ in (f).}
\vspace{-10pt}
\label{fig:1}
\end{figure}

Figs.~\ref{fig:1}(a,b) show experimental NS data 
for $\YbTiO$ at T $=9.1$\ K and  T $=1.4$\ K in the (hkk) plane.  
Sharp intense Bragg peaks
at integer reciprocal lattice positions were removed from the data to 
expose significant structure in the diffuse magnetic scattering.
Fig.~\ref{fig:1}a shows the NS map at T $=9.1$\ K.
 The magnetic correlations weaken with
increasing temperature. Indeed, most of the features present at T $=1.4$\ K
(Fig.~\ref{fig:1}b) are absent at $9.1$\ K, with only a weakened rod 
of scattering along $[111]$ and a
feature in the upper right 
corner at $3.5,2.25,2.25$ remaining. 
Fig.~\ref{fig:1}b shows the NS map at 
T $=1.4$ K, where the most interesting feature is the aforementioned
rod of scattering along the $[111]$ direction \cite{111rod,Ross}.  
Figure \ref{fig:1}b exhibits other features of interest
such as a weaker rod of scattering 
going from $400$ to $222$
  and intensity near the point 022.
The intensity of the feature centered on
$3.5,2.25,2.25$ does not change with temperature
(Fig.~\ref{fig:1}f) indicating that 
it is not magnetic in origin  and can therefore be omitted from
further consideration. 


To explain the NS pattern,  
we propose a Hamiltonian, ${H} = H_{\rm cf} + H_{\rm int}$,
that includes a crystal field (CF) part, $H_{\rm cf}$, 
and spin-spin interactions,
$H_{\rm int}=H_{\rm dip} + H_{\rm ex}$. 
The form of $H_{\rm cf}$ is fixed by the symmetry of the Yb$^{3+}$
environment.  The two sets of CF parameters that we use have
been determined in Refs.~\cite{Hodges-YbTO-PRL,Cao}. 
The magnetic Yb$^{3+}$ ion has electronic configuration $^2$F$_{7/2}$,
hence ${\rm J}=7/2$ and Land\'{e} factor $g_{\rm J}=8/7$.
The nearest neighbor
distance between Yb$^{3+}$ ions is $r_{\rm nn}=(a/4){\sqrt {2}}$, where $a=10.026$ \AA{} is the size of the conventional cubit unit cell \cite{Gardner-YbTO}.
This fixes the strength of the coupling 
$D=\frac
{
 {\mu_{0}(g_{\rm J} \mu_{\rm B})^2}}
{  {4\pi {(r_{\rm nn})}^3}
}
 \approx 0.01848$ K 
of the long-range magnetostatic dipolar interaction,
$H_{\rm dip} = \frac{1}{2} \sum_{(i,a;j,b)} \frac{D (r_{\rm nn})^3}
{\mid  \mathbf{R}_{ij}^{ab}\mid^{3}} ( \mathbf{{J}}_{i}^{a} \cdot \mathbf{{J}}_{j}^{b} -
3 (\mathbf{{J}}_{i}^{a} \cdot \hat{\mathbf{R}}_{ij}^{ab} )
 (\mathbf{{J}}_{j}^{b} \cdot \hat{\mathbf{R}}_{ij}^{ab} ) )$. 
We also consider $H_{\rm ex}$
which contains all nearest neighbour exchange
interactions, ${\cal J}_e$, 
that respect lattice symmetries.
There are four such nearest neighbor interactions \cite{1742-6596-145-1-012032}:
$H_{\rm Ising}=-\mathcal{J}_{\rm Ising}
\sum_{<i,a;j,b>}\left(\mathbf{{J}}_{i}^{a}
\cdot\mathbf{\hat{z}}^{a}\right)\left(\mathbf{{J}}_{j}^{b}\cdot\mathbf{
\hat{z}}^{b}\right)$, 
which couples the local $\left[111\right]$~${\hat z}$ components of 
$\mathbf{J}$, 
$H_{\rm iso}=
-\mathcal{J}_{\rm iso}\sum_{<i,a;j,b>}
\mathbf{{J}}_{i}^{a} \cdot \mathbf{{J}}_{j}^{b}$,  the standard isotropic exchange, 
$H_{\rm pd} =-\mathcal{J}_{\rm pd}\sum_{<i,a;j,b>} ( \mathbf{{J}}_{i}^{a} 
\cdot \mathbf{{J}}_{j}^{b} - 3 (\mathbf{{J}}_{i}^{a} \cdot
\hat{\mathbf{R}}_{ij}^{ab} ) (\mathbf{{J}}_{j}^{b}
 \cdot \hat{\mathbf{R}}_{ij}^{ab} ) )$, a pseudo-dipolar  
interaction of exchange origin and not
part of $H_{\rm dip}$ and, finally, 
$H_{\rm DM}=-\mathcal{J}_{\rm DM}\sum_{<i,a;j,b>}\boldsymbol{\Omega}^{a,b}_{{\rm DM}} 
\cdot \left(\mathbf{{J}}_{i}^{a} \times \mathbf{{J}}_{j}^{b}\right)$, 
the Dzyaloshinskii-Moriya (DM) interaction~\cite{Canals2}. 
 In all of these terms, 
$\mathbf{J}_{i}^{a}$
 denotes the angular momentum of the Yb$^{3+}$
 located at lattice $\mathbf{R}_{i}^{a}$
 (FCC lattice site $i$, 
 and tetrahedral sub-lattice site $a$) 
\cite{Gingras-MFT}
 and $\hat{\mathbf{R}}_{ij}^{ab}$ is a 
unit vector directed along $\mathbf{R}_{j}^{b}-\mathbf{R}_{i}^{a}$.
The relationship between $H_{\rm int}$ and a corresponding effective spin-1/2 model is discussed in Ref.~\cite{supmat}.

We use ${H}$ to compute the diffuse NS pattern within the RPA \cite{REM,PhysRevB.68.172407} (see Ref.~\cite{supmat} for justification of the usage of RPA here).
We first compute the single ion susceptibility, $\chi^{(0)}$,
 from $H_{\rm cf}$:
\begin{eqnarray}
\chi^{(0),\alpha\beta}_{a}(\omega)&=&\sum_{\mu,\nu}^{\Emu\ne\Enu}\frac{M^\alpha_{\nu\mu,a}
  M^\beta_{\mu\nu,a}}{\Emu-\Enu-\hbar(\omega+i 0^+)}(n_{\nu}-n_{\mu}) \nonumber\\
& &+\;\frac{\delta(\omega)}{k_B T}\sum_{\mu,\nu}^{\Emu=\Enu}
M^\alpha_{\nu\mu,a}M^\beta_{\mu\nu,a}n_{\nu},
\label{eq:chi0}
\end{eqnarray}
where $n_{\nu}$ is the thermal occupation fraction for CF state $\nu$. 
$M^\alpha_{\nu\mu,a}=\sum_{\bar{\alpha}}\langle \nu| {\rm J}^{\bar{\alpha}} | \mu
\rangle u^{\alpha}_{\bar{\alpha},a}$, 
where $u^{\alpha}_{\bar{\alpha},a}$ is the rotation matrix from the local ($\bar{\alpha}$) 
frame defined on sublattice $a$ to the global ($\alpha$) frame. 
The operator ${\mathbf{J}}^{\bar{\alpha}}$ 
acts on the CF states defined in the local $\bar\alpha$ quantization frame.
The CF wavefunctions $\vert \nu\rangle$, at sublattice site $a$, 
are obtained by diagonalizing $H_{\rm cf}$~\cite{Hodges-YbTO-PRL,Cao}. 
The interacting RPA susceptibility, $\chi(\mathbf{q},\omega)$ \cite{PhysRevB.68.172407}, 
is then
$
\chi^{\alpha\beta}_{ab}(\mathbf{q},\omega)+\sum_{\gamma,\delta,c}
\chi^{0,\alpha\gamma}_{a}(\omega)
\mathcal{J}^{\gamma\delta}_{ac}(\mathbf{q})\chi^{\delta\beta}_{cb}(\mathbf{q},\omega)=
\delta^{\ }_{ab}\chi^{0,\alpha\beta}_{a}(\omega),
$
where $\mathcal{J}(\mathbf{q})$ is the Fourier transformation of the
interaction matrix $\mathcal{J}(i,j)$ for
 Hamiltonian $H_{\rm int}=
-(1/2)\sum_{i,j;a,b;\alpha,\beta} 
{\rm J}_{i,a}^{\alpha}
\mathcal{J}_{ab}^{\alpha\beta}(i,j) 
{\rm J}_{j,b}^{\beta}$. The infinite lattice sum of the dipolar
interaction is computed using Ewald summation~\cite{Gingras-MFT}.   We solve for
$\chi^{\alpha\beta}_{ab}(\mathbf{q},\omega)$ numerically.
Finally, the NS  function, $S(\mathbf{q},\omega)$ \cite{REM,PhysRevB.68.172407}, 
 is given by
\begin{eqnarray}
\label{eqn:9}
S\left(\mathbf{q},\omega\right) & \propto & 
\frac{\mid f\left(\mathbf{Q}\right) \mid ^{2}}{k_{\rm B}\rm{T}} 
\displaystyle\sum_{\alpha,\beta} \displaystyle\sum_{a,b} \left(\delta_{\alpha\beta}  - \hat{Q}_{\alpha} \hat{Q}_{\beta} \right) \\
 \mbox{} 
& \times &
 \exp\left({-i \left(\mathbf{r}^{a} - \mathbf{r}^{b}\right)
  \cdot \mathbf{G}}\right) \rm{Re}\left(\chi_{ab}^{\alpha
  \beta}\left(\mathbf{q},\omega  \right)\right)
\nonumber
\end{eqnarray}
where $f\left(\mathbf{Q}\right)$ is the magnetic form factor for the Yb$^{3+}$ ion \cite{formfac}.
$\mathbf{Q} = \mathbf{q} + \mathbf{G}$ is the scattering wave vector where
$\mathbf{q}$ is a wave vector inside the first Brillouin zone, and $\mathbf{G}$ is an FCC
reciprocal lattice vector.  $\mathbf{r}^{a}$ and $\mathbf{r}^{b}$ are basis vectors for the tetrahedral
sublattice~\cite{Gingras-MFT}.


The fit to the experimental data was performed by computing the RPA scattering 
intensity along the measured lines in $\mathbf{Q}$
space (dashed lines in Fig. \ref{fig:1}b). The computed intensities were rescaled to the experimental count rate 
using the relation $S'\left(\mathbf{q}\right) =
c_{0}S\left(\mathbf{q}\right) + c_{1} +c_{2}\mid \mathbf{Q} \mid$ \cite{Yavorskii-PRL}, 
where the parameters $c_{0}$, $c_{1}$, and $c_{2}$ are the
same for all $\mathbf{Q}$ points.
 There are therefore seven adjustable parameters in total 
with the four exchange couplings, ${\cal J}_{e}$ and the three $c_n$ fitting parameters. 
The variance between the measured and calculated neutron data, 
with a contribution to the variance from fitting $\CW$ as well, was
minimized using a simulated annealing algorithm~\cite{Annealing}. 
As a first step, for simplicity and computational speed,
we make a static approximation~\cite{static_approx}
to $S\left(\mathbf{q}, \omega\right)$.
Within the moderate constraints \cite{supmat} 
of the procedure followed, the present calculation is suitable to reach the
main conclusion of this work: that significant anisotropic exchange
couplings ${\cal J}_e$ are necessary to account for the 
structure of the NS pattern of Yb$_2$Ti$_2$O$_7$.


Fig.~\ref{fig:1}d shows the RPA NS pattern in the $\left( {\rm h} {\rm k} {\rm k} \right)$ plane
at $1.4$ K obtained from simulated annealing fits to the experimental neutron
scattering map of Fig.~\ref{fig:1}b. 
The model data in Fig.~\ref{fig:1}d is obtained from a $H_{\rm ex}$ with 
$\mathcal{J}_{\rm Ising} = 0.81$ K, $\mathcal{J}_{\rm iso} = 0.22$ K,
$\mathcal{J}_{\rm pd} = -0.29$ K, and $\mathcal{J}_{\rm DM} = -0.27$ K.
The calculated intensity matches the experimental data (Fig.~\ref{fig:1}b)
well, providing strong evidence that our model $H_{\rm ex}$ contains the
correct interactions for $\YbTiO$.  A cut along $\left[3-{\rm k}, {\rm k}, {\rm k}\right]$
(Fig.~\ref{fig:1}c) emphasizes the quantitative agreement between the computed
and experimental NS intensities.  A similar quality of fit was obtained for
other line scans (dashed lines in Fig.~\ref{fig:1}b) 
shown in the supplemental material \cite{supmat}. We carried out the
fitting procedure using CF parameters for $H_{\rm cf}$ taken from
Refs.~\cite{Hodges-YbTO-PRL,Cao} finding that the exchange couplings do
not change significantly. Fig.~\ref{fig:1}e and the dashed (black) 
line in Fig.~\ref{fig:1}c show the NS
intensity calculated for a model $H_{\rm int}$ with only long-range dipolar
and isotropic exchange interactions ($\mathcal{J}_{\rm Ising} =
\mathcal{J}_{\rm pd} = \mathcal{J}_{\rm DM} = 0$) with $\mathcal{J}_{\rm iso}
= 0.06$ K, determined by fitting $\CW$, and refitted $c_n$'s.
  Clearly this model and the resulting NS pattern do not describe the experimental data
(Fig.~\ref{fig:1}b) well at all.
Similarly, a Hamiltonian with isotropic-only 
exchange does not describe the local susceptibility,
 $\chi_{\rm local}$, well \cite{Cao,Bonville-JPC,Malkin,Thompson-JPCM}. 
On the other hand, the present anisotropic Hamiltonian describes 
$\chi_{\rm local}$ with no adjustable parameters \cite{Thompson-JPCM}.

\begin{figure}[ht]
\begin{center}
\includegraphics[width=0.5\columnwidth]{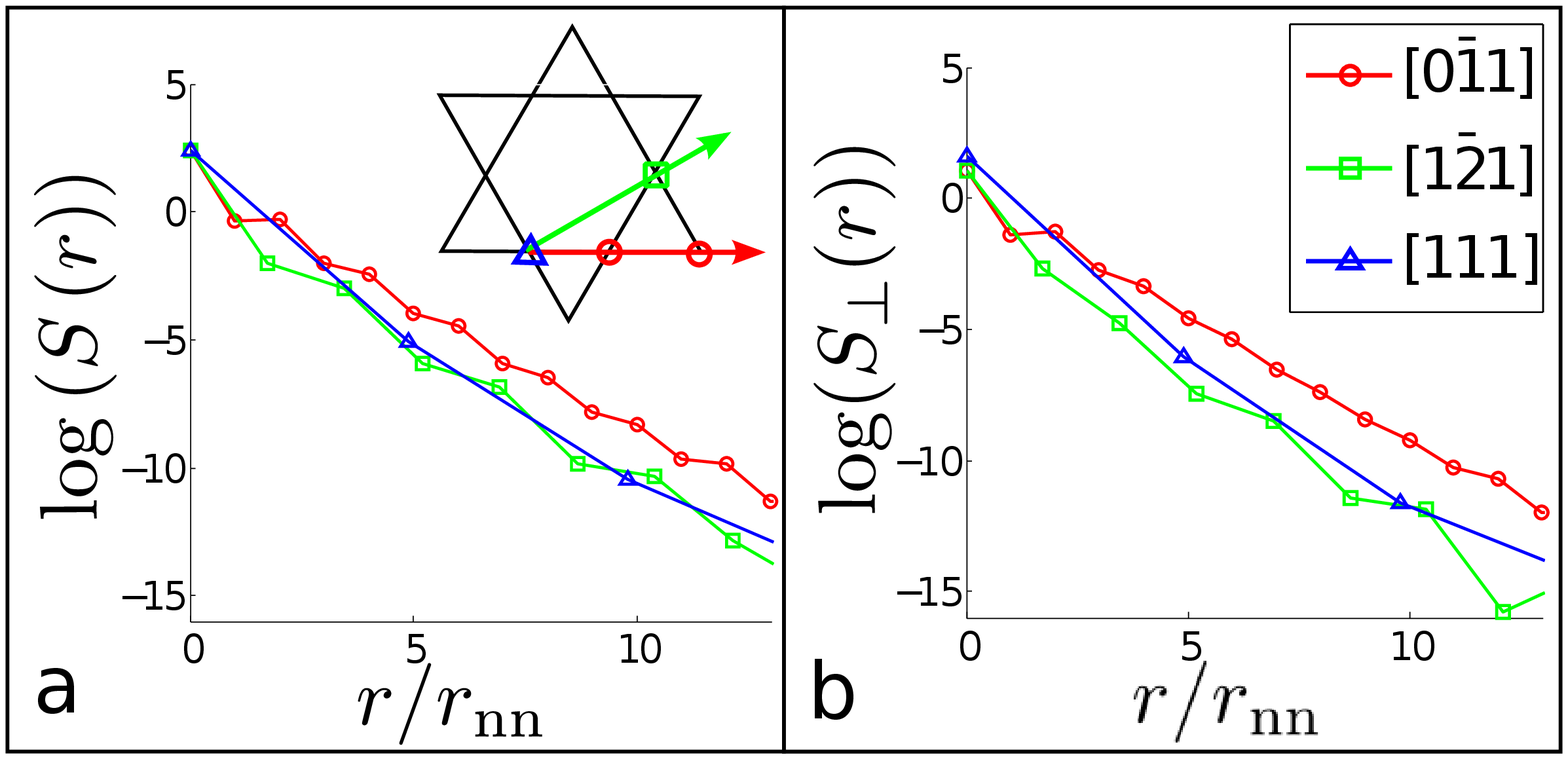}
\caption{(Color online) Real space
spin-spin correlation
 functions (a) $S(\mathbf{r})$ and
 (b) $S_\perp(\mathbf{r})$ (see text) computed at T $=1.4$\ K
  using the anisotropic exchange model and plotted along various
  crystallographic directions.
  }
\vspace{-20pt}
\label{fig:4}
\end{center}
\end{figure}

To rationalize the direct space origin of rods of NS intensity, we computed the spin-spin correlation function from the reciprocal space RPA susceptibility
$\langle J^{\alpha}_{a}\left(\mathbf{r}\right) 
    J^{\beta}_{b}\left( \mathbf{0}\right)\rangle = k_{\rm
  B}T\int\chi^{\alpha\beta}_{ab}\left(\mathbf{q}\right)\exp
  \left(\mathbf{q}\cdot\mathbf{r}\right)d\mathbf{q} 
$
where $\mathbf{r} = \mathbf{R}_{j}^{b}-\mathbf{R}_{i}^{a}$. The integral was performed numerically over the first Brillouin zone. 
We considered the isotropic real space correlations (by summing over all directions $\alpha$ in spin space) 
$S(\mathbf{r})\equiv \sum_{\alpha} \langle J^{\alpha}_{a}\left(\mathbf{r}\right) J^{\alpha}_{b}\left( \mathbf{0}\right)\rangle$, and also $S_\perp(\mathbf{r}) \equiv k_{\rm B}T\int [\delta_{\alpha\beta}- \hat{q}_{\alpha}\hat{q}_{\beta}] \chi^{\alpha\beta}_{ab}\left(\mathbf{q}\right)\exp \left(\mathbf{q}\cdot\mathbf{r}\right)d\mathbf{q}$ whose Fourier transform is measured in the NS.  
Figure \ref{fig:4} shows $S(\mathbf{r})$ and $S_{\perp}(\mathbf{r})$ for $\mathbf{r}$ taken along the
  $[111]$ direction, and $[0\bar{1}1]$, $[1\bar{2}1]$ perpendicular to $[111]$. $S(\mathbf{r})$ and $S_{\perp}(\mathbf{r})$ are similar,
 so we discuss both together. The correlation lengths for these three directions were extracted, assuming exponential decay, and found not to differ greatly within our margin of error. However, the $[0\bar{1}1]$ correlations are larger than those of the other two directions indicating some degree of spontaneous decoupling of the kagome planes albeit perhaps with a quasi-isotropic correlation length. Since the $[0\bar{1}1]$ directions lie within two kagome planes and the $[1\bar{2}1]$ directions lie in only one, truly two dimensional correlations may have been
be expected to lead to the splitting $S(\mathbf{r})_{[0\bar{1}1]}>S(\mathbf{r})_{[1\bar{2}1]}>S(\mathbf{r})_{[111]}$. This is not borne out by our results $-$
 the similarity of $S(\mathbf{r})$ for the $[1\bar{2}1]$ and $[111]$ directions indicates that the correlations are strongest along spin chains. Whereas the
NS intensity in the (hhk) plane, considered on its own, 
suggests that the correlations are quasi two-dimensional \cite{Ross},
 with a weak decoupling of the kagome planes, the real space correlations
 within our model do not support this simple picture.

From the determined $H_{\rm ex}$, RPA predicts a second order phase transition
to a ferromagnetic phase (ordering wavevector ${\mathbf q} = 0$) at a critical
temperature $T_c^{\rm RPA} \approx 1.2$ K.  We expect thermal and quantum fluctuations to renormalize the values
of the anisotropic exchange ${\cal J}_e$ determined above.

In summary, we have presented diffuse neutron scattering (NS) 
maps of Yb$_{2}$Ti$_{2}$O$_{7}$ in its paramagnetic regime,
 finding
rods of scattering in the $\langle 111\rangle$ directions. By fitting this data to
NS computed from a candidate model Hamiltonian, we found a
set of couplings that reproduce the main features of this NS
pattern. 
The Hamiltonian includes, as an essential component, sizeable anisotropic
exchange interactions.  This suggests that anisotropic exchange might be important in other
{\it A}$_{2}${\it B}$_{2}${\rm O}$_{7}$ rare earth magnets \cite{Cao,Malkin}.
We find that the rods of scattering occur without any symmetry breaking from, for
example, a structural phase transition. We anticipate that our results will
allow for a greater understanding of the nature of the phase transition at
$240$ mK and of the low temperature phase of $\YbTiO$.


We thank Y.-J. Kao for useful discussions and contributions at the earliest stage of this project.  We acknowledge useful discussions with B. Gaulin, K. Ross and J. Ruff.  This work was funded by the NSERC of Canada and the CRC Program (M.G., Tier 1).


\section{Supplemental Material}

This document is a supplement to our main article, where we explain the relationship between our model Hamiltonian and effective spin-$1/2$ models appropriate for this material, such as the model considered in Ref.~\cite{Onoda1} for Pr$_2M_2$O$_7$ ($M=$~Ir, Zr, Sn) pyrochlore materials. We also provide details on the validity of the random phase approximation (RPA) for computing the neutron scattering pattern at $T=1.4$ K, constraints on the results of our RPA calculations, a short discussion of energy integrated neutron scattering results, and additional line scans to demonstrate the success of our model.

\section {Effective Spin Half Equivalent Hamiltonian}

In Yb$_2$Ti$_2$O$_7$, the energy gap separating the crystal field ground state and the first excited doublet is $\Delta \sim 620$ K, very large compared to the magnetic dipolar and exchange (bilinear or higher multipolar \cite{Santini-RMP}) interactions at play in the compound. Hence, as in spin ices \cite{Gingras-Springer} and LiHoF$_4$ \cite{Tabei-LiHoF4}, but unlike the Tb$_2$Ti$_2$O$_7$ \cite{Molavian} and Tb$_2$Sn$_2$O$_7$ \cite{McClarty} pyrochlores, it is safe to project the microscopic Hamiltonian, $H_{\rm int}$, describing the interactions between Yb$^{3+}$ ions, into the Hilbert space spanned solely by a direct product of the states in the Yb$^{3+}$ crystal field ground state doublet.
Such a projection allows us to rewrite $H_{\rm int}$, irrespective of its original complicated (even possibly multipolar) form, in terms of an effective Hamiltonian, $H_{\rm eff}$, expressed solely in terms of anisotropic couplings between effective spin-1/2 operators.
In principle, we could have considered such an effective model in the main body of our paper, parametrized by effective couplings $\tilde {\cal J}_{\rm iso}$, $\tilde {\cal J}_{\rm Ising}$, $\tilde {\cal J}_{\rm pd}$ and $\tilde {\cal J}_{\rm DM}$, and expressed the neutron scattering function in terms of components of ${\mathbf J}$ projected onto the crystal field ground state doublet.  Again, this is possible since there is negligible operator correction to the two-point correlation function that enters in the scattering function because $\Delta$ is so large compared to $H_{\rm int}$.
 
In this context, it is therefore important to view the model describing the exchange interactions, $H_{\rm ex}$, with couplings ${\cal J}_{\rm iso}$, $ {\cal J}_{\rm Ising}$, $ {\cal J}_{\rm pd}$ and $ {\cal J}_{\rm DM}$, {\it not} as a microscopic model capturing the correct physics on energy scales comparable to the crystal field splitting (which may include multipolar interactions) \cite{Santini-RMP}, but rather one whose projection gives the correct effective low-energy theory in terms of pseudo-spin-1/2. Our model is therefore an ``un-projected'' version of the effective spin-1/2 Hamiltonian $H_{\rm eff}$ describing Yb$_2$Ti$_2$O$_7$, which we have taken to be a strictly bilinear Hamiltonian, ${\cal J}_{ij}^{uv} {\rm J_i^u}{\rm J_j^v}$. This was done for presentation sake and to relate our approach to what was used previously in a random phase approximation treatment of Tb$_2$Ti$_2$O$_7$ \cite{PhysRevB.68.172407}. In the rest of this section, we relate our bilinear couplings $\{{\cal J}_{\rm e}\}$ to the anisotropic couplings between the components of a spin-1/2 in an effective $H_{\rm eff}$ Hamiltonian.

The exchange model $H_{\rm ex}$ in our work consists of bilinear exchange terms between the full blown angular momentum operators ${\mathbf J}$ for the Yb$^{3+}$ ions.  As stated above, the exchange model can be written in terms of an effective spin-$1/2$ model.  This is done by computing the expectation value of the three components of ${\mathbf J}$ within the ground state doublet and defining $g_\parallel = 2g_J\langle \psi_\pm \mid {\mathbf J^z} \mid \psi_\pm\rangle$ and $g_\perp = 2g_J\langle \psi_\pm \mid {\mathbf J^x} \mid \psi_\pm\rangle = 2g_J\langle \psi_\pm \mid {\mathbf J^y} \mid \psi_\pm\rangle$. Here $\mid \psi_\pm \rangle$ are the two states that make up the ground state doublet of the crystal field, ${\mathbf J^a}$ is expressed in the local coordinate system for each corner of a tetrahedron, $g_J = 8/7$ is the Land\'{e} factor, and $\hat{z}$ is the corresponding $[111]$ cubic diagonal direction.  In this coordinate system, $g_\parallel$ and $g_\perp$ are the only non-zero terms, and we can construct the $\mathbf g$ tensor:
\begin{equation}
{\mathbf g} = \left( \begin{array}{ccc}
g_\perp & 0 & 0 \\
0 & g_\perp & 0 \\
0 & 0 & g_\parallel \end{array} \right) \; .
\end{equation}
For the crystal field of Ref.~\cite{Hodges-YbTO-JPC} (Ref.~\cite{Bonville-JPC}), $g_\parallel = 1.77$ ($g_\parallel = 2.25$) and $g_\perp = 4.18$ ($g_\perp = 4.1$).
Using this $\mathbf g$ tensor, the relationship between the full blown angular momentum operators and the effective spin-$1/2$ operators is given by ${\mathbf J} = \tfrac{{\mathbf g}}{g_J}\vec{S}_{\rm eff} = \tfrac{{\mathbf g}}{2g_J}\vec{\sigma}$, where $\vec{S}_{\rm eff}$ is an effective spin-$1/2$ and $\vec{\sigma}$ is a vector of Pauli matrices.

Using this relationship, we can recast $H_{\rm ex}$ in terms of effective spin-$1/2$ operators by replacing all of the terms ${\mathbf J}_i^a$ with $\tfrac{{\mathbf g}_i^a}{2g_J}\vec{\sigma}_i^a$ where the labels $i$ and $a$ are required because we are now passing from a local coordinate system to the global cartesian coordinate system, meaning that ${\mathbf g}_i^a$ now changes from site to site.  Performing this transformation, we obtain $H_{\rm eff} = H'_{\rm Ising} + H'_{\rm iso} + H'_{\rm pd} + H'_{\rm DM}$, where:
\begin{align} 
H'_{\rm Ising}&=-\frac{\mathcal{J}_{\rm Ising}}{4g_J^2}\sum_{<i,a;j,b>}\left({{\mathbf g}_i^a}{{\vec{\sigma}}}_{i}^{a}\cdot\mathbf{\hat{z}}^{a}\right) \left({\mathbf g}_j^b{{\vec{\sigma}}}_{j}^{b}\cdot\mathbf{\hat{z}}^{b}\right),\\
H'_{\rm iso}&= -\frac{\mathcal{J}_{\rm iso}}{4g_J^2}\sum_{<i,a;j,b>}{{\mathbf g}_i^a}{{\vec{\sigma}}}_{i}^{a} \cdot {{\mathbf g}_j^b}{{\vec{\sigma}}}_{j}^{b},\\
H'_{\rm pd} &=-\frac{\mathcal{J}_{\rm pd}}{4g_J^2}\sum_{<i,a;j,b>} ({{\mathbf g}_i^a}{{\vec{\sigma}}}_{i}^{a} \cdot{{\mathbf g}_j^b}{{\vec{\sigma}}}_{j}^{b} - 3 ({{\mathbf g}_i^a}{{\vec{\sigma}}}_{i}^{a} \cdot \hat{\mathbf{R}}_{ij}^{ab} ) ({{\mathbf g}_j^b}{{\vec{\sigma}}}_{j}^{b} \cdot \hat{\mathbf{R}}_{ij}^{ab} ) ),\\
H'_{\rm DM}&=-\frac{\mathcal{J}_{\rm DM}}{4g_J^2}\sum_{<i,a;j,b>}\boldsymbol{\Omega}^{a,b}_{{\rm DM}} \cdot \left({{\mathbf g}_i^a}{{\vec{\sigma}}}_{i}^{a} \times {{\mathbf g}_j^b}{{\vec{\sigma}}}_{j}^{b}\right).
\end{align}

In addition to the representation of the symmetry allowed bilinear exchange interactions used in this paper, other representations are possible, which are linear combinations of the terms used in the original representation of $H_{\rm eff}$. One other representation is given by Ref.~\cite{Onoda2}:
\begin{eqnarray}
\label{eq:onoda}
  H_{\rm eff}&=&-J_{\rm nn}\sum^{\rm nn}_{\langle \bm{r},\bm{r}'\rangle}\left[
    g^\parallel\hat{\sigma}^z_{\bm{r}}\hat{\sigma}^z_{\bm{r}'}
    +g^\bot\left(\hat{\sigma}^x_{\bm{r}}\hat{\sigma}^x_{\bm{r}'}+\hat{\sigma}^y_{\bm{r}}\hat{\sigma}^y_{\bm{r}'}\right)
    \right.\nonumber\\
    &&\left.\ \ \ \ \ \ \ \ \ \ \ \ \ \ 
    +g^q\left(\left(\hat{\vec{\sigma}}_{\bm{r}}\cdot\vec{n}_{\bm{r},\bm{r}'}\right)\left(\hat{\vec{\sigma}}_{\bm{r}'}\cdot\vec{n}_{\bm{r},\bm{r}'}\right)
    -\left(\hat{\vec{\sigma}}_{\bm{r}}\cdot\vec{n}'_{\bm{r},\bm{r}'}\right)\left(\hat{\vec{\sigma}}_{\bm{r}'}\cdot\vec{n}'_{\bm{r},\bm{r}'}\right)\right)
    \right.\nonumber\\
    &&\left.\ \ \ \ \ \ \ \ \ \ \ \ \ \ 
    +g^K\left(\hat{\sigma}^z_{\bm{r}}\left(\hat{\vec{\sigma}}_{\bm{r}'}\cdot\vec{n}_{\bm{r},\bm{r}'}\right)+\left(\hat{\vec{\sigma}}_{\bm{r}}\cdot\vec{n}_{\bm{r},\bm{r}'}\right)\hat{\sigma}^z_{\bm{r}'}\right)
    \right],
\end{eqnarray}
In this representation $x$, $y$, and $z$ refer to the local coordinates at each corner of the tetrahedra, $\hat{\vec{\sigma}}_{\bm{r}} = \left(\hat{\sigma}^x_{\bm{r}},\hat{\sigma}^y_{\bm{r}}\right)$, $\vec{n}_{\bm{r},\bm{r}'} = \left(\cos \phi_{\bm{r},\bm{r}'}, -\sin \phi_{\bm{r},\bm{r}'}\right)$, $\vec{n}'_{\bm{r},\bm{r}'} = \left(\sin \phi_{\bm{r},\bm{r}'},\cos \phi_{\bm{r},\bm{r}'}\right)$, and $\phi_{\bm{r},\bm{r}'} = 0,2\pi/3,-2\pi/3$ \cite{Onoda2}.  The various $g^\alpha$ terms are defined as:
\begin{align}
 g^\parallel&=1-8\sqrt{6}x-\frac{9}{2}x^2-3\sqrt{6}x^3+\frac{63}{16}x^4,
  \label{eq:g^parallel}\\
  g^\bot&=1+4\sqrt{6}x+\frac{45}{2}x^2-3\sqrt{6}x^3+\frac{9}{16}x^4,
  \label{eq:g^bot}\\
  g^q&=-2\left(1-2\sqrt{6}x+9x^2-3\sqrt{6}x^3+\frac{9}{4}x^4\right),
  \label{eq:g^q}\\
  g^K&=2\sqrt{2}\left(1+\sqrt{6}x-\frac{45}{4}x^2+\frac{15}{4}\sqrt{6}x^3-\frac{9}{8}x^4\right),
  \label{eq:g^K}
\end{align}
where  $x=V_{pf\pi}/V_{pf\sigma}$, is the ratio of two Slater-Koster parameters, representing transfer integrals between $p_x/p_y$ and $f_{x\left(5z^2-r^2\right)}/f_{y\left(5z^2-r^2\right)}$ orbitals and $p_z$ and $f_{\left(5z^2-3r^2\right)z}$ orbitals, respectively \cite{Onoda2}.
This representation can be related to our original notation by using the relationships between global cartesian coordinates and the local coordinate system for each corner of the tetrahedral sublattice, and the relationships between the exchange terms in the body of the paper and the exchange terms in Eq.~\ref{eq:onoda}.  The net result of these relationships is expressed by the following relations:
\begin{align}
-J_{\rm nn}g^\parallel &= -\frac{g_\parallel^2}{12g_J^2}\left(-3\mathcal{J}_{\rm Ising} + \mathcal{J}_{\rm iso} -5\mathcal{J}_{\rm pd} - 4\mathcal{J}_{\rm DM}\right),\\
-J_{\rm nn}g^\perp &= -\frac{g_\perp^2}{48g_J^2}\left(\mathcal{J}_{\rm iso} -\frac{1}{2}\mathcal{J}_{\rm pd} + \frac{1}{2}\mathcal{J}_{\rm DM}\right),\\
-J_{\rm nn}g^q &= \frac{g_\perp^2}{24g_J^2}\left(\mathcal{J}_{\rm iso} + \frac{7}{4}\mathcal{J}_{\rm pd} - \mathcal{J}_{\rm DM}\right),\\
-J_{\rm nn}g^K &= -\frac{\sqrt{2} g_\parallel g_\perp}{12g_J^2}\left(\mathcal{J}_{\rm iso} - \frac{1}{2}\mathcal{J}_{\rm pd} + 2\mathcal{J}_{\rm DM}\right).
\end{align}

\begin{figure}[htbp]
\begin{center}
\includegraphics[scale=0.6]{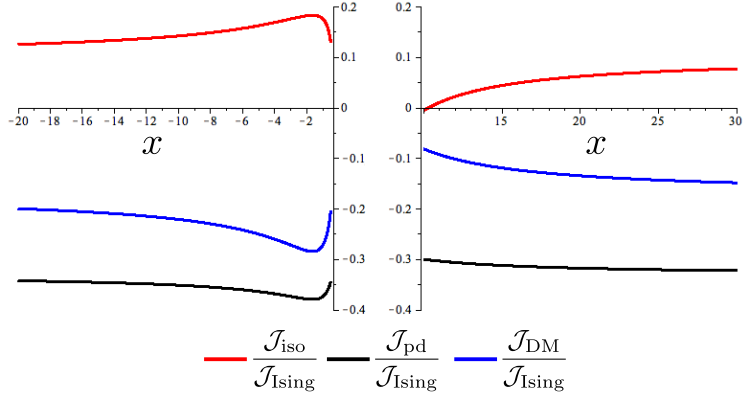}
\caption{(Color online). Plots of the ratios $\frac{\mathcal{J}_{\rm iso}}{\mathcal{J}_{\rm Ising}}$, $\frac{\mathcal{J}_{\rm pd}}{\mathcal{J}_{\rm Ising}}$, and $\frac{\mathcal{J}_{\rm DM}}{\mathcal{J}_{\rm Ising}}$ as a function of $x$ computed from the work of Ref.~\cite{Onoda2} for the ranges $x\approx\left[-20,-0.5\right]$ and $x\approx\left[10,30\right]$.}
\label{fig:S:2}
\end{center}
\end{figure}

By inverting these relations, we can compute the ratios $\frac{\mathcal{J}_{\rm iso}}{\mathcal{J}_{\rm Ising}}$, $\frac{\mathcal{J}_{\rm pd}}{\mathcal{J}_{\rm DM}}$, and $\frac{\mathcal{J}_{\rm DM}}{\mathcal{J}_{\rm Ising}}$, from the work of Ref.~\cite{Onoda2}, finding that, for $x\approx\left[-\infty,-0.5\right]$ and $x\approx\left[20,\infty\right]$, there is fair agreement in terms of sign and magnitude with those of our work $\frac{\mathcal{J}_{\rm iso}}{\mathcal{J}_{\rm Ising}} = 0.28$, $\frac{\mathcal{J}_{\rm pd}}{\mathcal{J}_{\rm Ising}} = -0.36$, and $\frac{\mathcal{J}_{\rm DM}}{\mathcal{J}_{\rm Ising}} = -0.33$, see Fig. \ref{fig:S:2}. In the range $x\approx\left[-0.5,10\right]$, the ratios $\frac{\mathcal{J}_{\rm iso}}{\mathcal{J}_{\rm Ising}}$, $\frac{\mathcal{J}_{\rm pd}}{\mathcal{J}_{\rm DM}}$, and $\frac{\mathcal{J}_{\rm DM}}{\mathcal{J}_{\rm Ising}}$ diverge when $\mathcal{J}_{\rm Ising}\rightarrow 0$,Êand fluctuate rapidly, making any comparison to our model very difficult. 

In addition to these effective spin-$1/2$ models, we can relate the notation used in this work, $H_{\rm ex} \equiv  H_{\rm Ising} + H_{\rm iso} + H_{\rm pd} + H_{\rm DM}$, to the form of the nearest neighbour invariants described in Ref.~\cite{1742-6596-145-1-012032} as $\left\{\chi\right\}$.  These are:
\begin{eqnarray}
\chi_{1} & = & {\rm J}_{1}^{x}{\rm J}_{2}^{x} + {\rm J}_{1}^{y}{\rm J}_{2}^{y} +{\rm  J}_{1}^{x}{\rm J}_{3}^{x} +{\rm  J}_{1}^{z}{\rm J}_{3}^{z} + {\rm J}_{1}^{y}{\rm J}_{4}^{y} + {\rm J}_{1}^{z}{\rm J}_{4}^{z}\\
 & & + {\rm J}_{2}^{y}{\rm J}_{3}^{y} + {\rm J}_{2}^{z}{\rm J}_{3}^{z} + {\rm J}_{2}^{x}{\rm J}_{4}^{x} + {\rm J}_{2}^{z}{\rm J}_{4}^{z} + {\rm J}_{3}^{x}{\rm J}_{4}^{x} + {\rm J}_{3}^{y}{\rm J}_{4}^{y}, \nonumber\\
\chi_{2} & = & {\rm J}_{1}^{z}{\rm J}_{2}^{z} + {\rm J}_{1}^{y}{\rm J}_{3}^{y} + {\rm J}_{1}^{x}{\rm J}_{4}^{x} + {\rm J}_{2}^{x}{\rm J}_{3}^{x} + {\rm J}_{2}^{y}{\rm J}_{4}^{y} + {\rm J}_{3}^{z}{\rm J}_{4}^{z},\\
\chi_{3} & = & {\rm J}_{1}^{x}{\rm J}_{2}^{y} + {\rm J}_{1}^{y}{\rm J}_{2}^{x} + {\rm J}_{1}^{x}{\rm J}_{3}^{z} + {\rm J}_{1}^{z}{\rm J}_{3}^{x} + {\rm J}_{1}^{y}{\rm J}_{4}^{z} + {\rm J}_{1}^{z}{\rm J}_{4}^{y}\\
 & & - {\rm J}_{2}^{y}{\rm J}_{3}^{z} - {\rm J}_{2}^{z}{\rm J}_{3}^{y} - {\rm J}_{2}^{x}{\rm J}_{4}^{z} - {\rm J}_{2}^{z}{\rm J}_{4}^{x} - {\rm J}_{3}^{x}{\rm J}_{4}^{y} - {\rm J}_{3}^{y}{\rm J}_{4}^{x}, \nonumber\\
\chi_{4} & = & \left[{\rm J}_{1}^{x},{\rm J}_{2}^{z}\right] + \left[{\rm J}_{1}^{y},{\rm J}_{2}^{z}\right] + \left[{\rm J}_{1}^{x},{\rm J}_{3}^{y}\right] + \left[{\rm J}_{1}^{z},{\rm J}_{3}^{y}\right] + \left[{\rm J}_{1}^{y},{\rm J}_{4}^{x}\right] + \left[{\rm J}_{1}^{z},{\rm J}_{4}^{x}\right]\\
 & & + \left[{\rm J}_{2}^{x},{\rm J}_{3}^{z}\right] + \left[{\rm J}_{2}^{y},{\rm J}_{3}^{x}\right] + \left[{\rm J}_{2}^{x},{\rm J}_{4}^{y}\right] + \left[{\rm J}_{2}^{y},{\rm J}_{4}^{z}\right] + \left[{\rm J}_{3}^{z},{\rm J}_{4}^{y}\right] + \left[{\rm J}_{3}^{x},{\rm J}_{4}^{z}\right], \nonumber
\end{eqnarray}
where $\left[J_i^u,J_j^v\right]\equiv J_i^uJ_j^v - J_i^vJ_j^u$.  In this notation, we may write $H_{\rm ex} = -J_1\chi_1 - J_2\chi_2 - J_3\chi_3 - J_4\chi_4$.  The relationships between the $\left\{{\mathcal J}_e\right\}$ and $\left\{J_n\right\}$ are given by:
\begin{eqnarray}
\mathcal{J}_{\rm Ising} & = & -3J_1 + 3J_2 + 3J_3 \\
\mathcal{J}_{\rm iso} & = & \frac{1}{3}J_1 + \frac{2}{3}J_2 + \frac{1}{3}J_3 \\
\mathcal{J}_{\rm pd} & = & \frac{2}{3}J_1 - \frac{2}{3}J_2 - \frac{4}{3}J_3 \\
\mathcal{J}_{\rm DM} & = & J_1 - J_2 - J_3 + J_4
\end{eqnarray}

\section{Justification for the use of the Random Phase Approximation}

The random phase approximation (RPA), which uses the bare single-ion anisotropy as non-interacting reference susceptibility, is well
justified in the case of $\YbTiO$ at $T=1.4$ K for two main reasons.
Firstly, $H_{\rm ex}$ and $H_{\rm dip}$ are weak ($\sim 10^0$ K and $\sim 0.02$ K, respectively) compared to the large ($\Delta\sim 620$ K \cite{Hodges-YbTO-JPC}) energy gap between the ground state crystal field doublet and the first excited crystal field doublet \cite{Hodges-YbTO-JPC}.  This implies that there is negligible interaction-induced admixing between the ground and first excited crystal field doublets, unlike the case of Tb$_2$Ti$_2$O$_7$, where admixing between the two lowest energy crystal field doublets is significant, due to the much smaller energy gap between between the ground and first excited states ($\sim 18$ K) \cite{Molavian}. Secondly, the correlations used to calculate the neutron scattering are at 1.4 K,
a temperature of about twice the Curie-Weiss temperature ($\CW\sim 0.75$ K \cite{Bramwell-JPC,Hodges-YbTO-JPC}), with the system
in the paramagnetic regime, where RPA should be reasonably valid.



\section{Constraints of the RPA Method and Static Approximation}

For the set of parameters, $\mathcal{J}_e\equiv \{{\mathcal J}_{\rm Ising},{\mathcal J}_{\rm iso},{\mathcal J}_{\rm pd},{\mathcal J}_{\rm DM} \}$, obtained from our minimization procedure using the static approximation, we carried out a more rigorous energy integrated calculation of the full dynamical scattering function, $S\left(\mathbf{q},\omega\right)$. Performing this calculation, we obtain a very similar, but not quantitatively identical, reciprocal space map of intensities to the scattering pattern obtained using the static approximation.

When using the RPA, it is also important to consider whether or not correlation effects, that develop near phase transitions, are being neglected.  In this work we use the RPA to
perform calculations at $T=1.4$ K, approximately 5 times the reported transition temperature of $\YbTiO$ ($\sim 240$ mK \cite{Hodges-YbTO-JPC}), and twice the Curie-Weiss temperature ($\CW\sim 0.75$ K \cite{Bramwell-JPC,Hodges-YbTO-JPC}).
Due to these correlation effects, we expect the fitted anisotropic exchange coupling to be renormalized as a function of the temperature at which the fit is computed.  The change should be of order $1/\left(T-\CW\right)^2 \approx 25$\%, the leading correction in a high temperature expansion of $\chi\left({\mathbf q}\right)$, compared to that of the RPA approximation of $\chi\left({\mathbf q}\right)$.
Despite these constraints, each unique set of couplings will generate a distinct diffuse scattering pattern that should be captured well by the RPA calculations to within a scaling factor that will differ from the scaling factor determined from the RPA calculations by a factor of order one. This is based on the fact that the physical scale is fixed by the Curie-Weiss temperature, and we included this as a constraint when performing the simulated annealing fit.


\section{Additional Neutron Scattering Line Scans}

\begin{figure}[htbp]
\begin{center}
\includegraphics[scale=0.75]{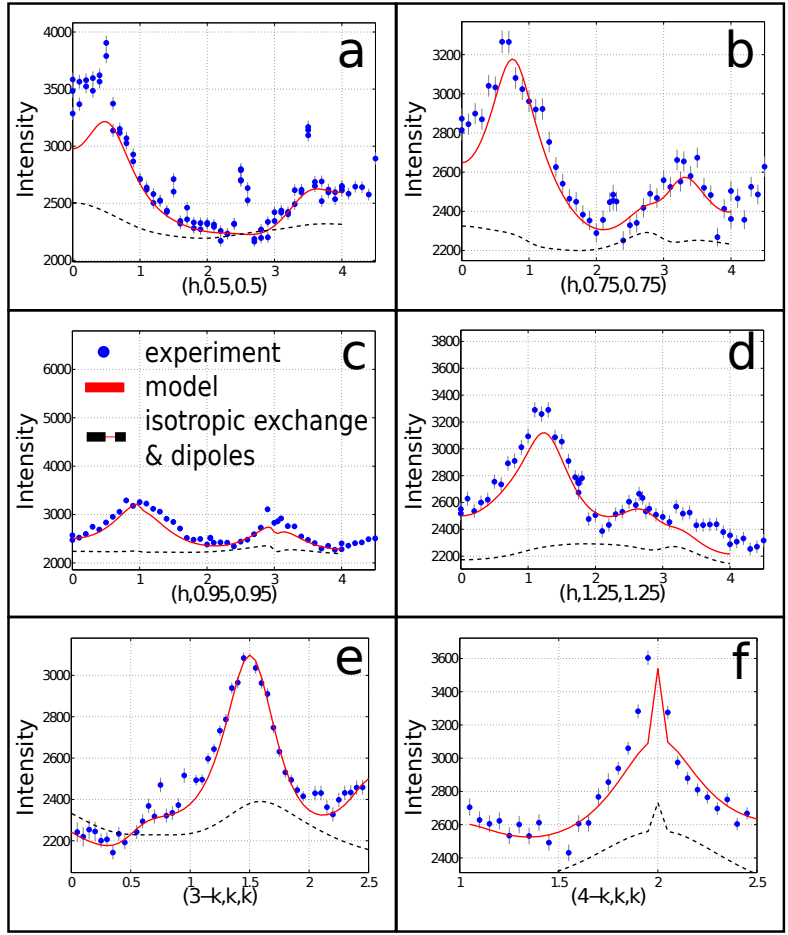}
\caption{(Color online).
Panels {\bf a} through {\bf f} show linescans within the $[hkk]$ plane.  The blue dots are experimental diffuse neutron scattering data taken at $T=1.4$ K.  The solid red lines are calculated from the {\it single set} of anisotropic exchange couplings
(${\mathcal J}_{\rm iso},
 {\mathcal J}_{\rm Ising},
 {\mathcal J}_{\rm pd},
 {\mathcal J}_{\rm DM}$)
presented in
the body of the paper,
 and the black dashed line is the prediction of a model with isotropic exchange and dipolar interactions, with the scale of the isotropic exchange set solely by the value of the Curie-Weiss temperature ($\CW \sim 0.75$ K \cite{Bramwell-JPC,Hodges-YbTO-JPC})}
\label{fig:S:1}
\end{center}
\end{figure}

To supplement the results presented in Fig.~\ref{fig:1} of the main body of the paper, Fig.~\ref{fig:S:1} shows additional cuts through the $[hkk]$ plane at $T=1.4$ K.  We chose this particular temperature in an effort to maximize the signal to noise ratio of the $[111]$ rod feature. Using data collected at higher temperatures, where the RPA would suffer less from temperature renormalization effects, would mean a loss of scattering intensity, and thus greater difficulty distinguishing the features of interest from background scattering.
Higher temperature data with high signal-to-noise ration would
be very desirable to ascertain further the quantitative accuracy and aforementioned temperature renormalization of the exchange parameters determined in the present work.

\newpage

\end{document}